# Terrestrial Implications of Cosmological Gamma-Ray Burst Models

S. E. Thorsett

*Joseph Henry Laboratories and Department of Physics,*
*Princeton University, Princeton, NJ 08544*

## Abstract

The observation by the BATSE instrument on the Compton Gamma Ray Observatory that gamma-ray bursts (GRBs) are distributed isotropically around the Earth but nonuniformly in distance has led to the widespread conclusion that GRBs are most likely to be at cosmological distances, making them the most luminous sources known in the Universe. If bursts arise from events that occur in normal galaxies, such as neutron star binary inspirals, then they will also occur in our Galaxy about every hundred thousand to million years. The gamma-ray flux at the Earth due to a Galactic GRB would far exceed that from even the largest solar flares. The absorption of this radiation in the atmosphere would substantially increase the stratospheric nitric oxide concentration through photodissociation of $N_2$, greatly reducing the ozone concentration for several years through $NO_x$ catalysis, with important biospheric effects due to increased solar ultraviolet flux. A nearby GRB may also leave traces in anomalous radionuclide abundances.

*Subject headings:* Gamma Rays: Bursts

The discovery that GRBs do not originate on nearby neutron stars has been the most surprising and important result from the BATSE burst experiment. The nearly one thousand bursts observed are remarkably isotropically distributed around the Earth, but BATSE is sampling bursts out to the "edge" of the distribution, as shown by a relative paucity of weak bursts (Fishman *et al.* 1994, Meegan *et al.* 1994). This is incompatible with a source population in the Galactic disk, where sampling distances large enough to exhibit inhomogeneity would also show a quadrupolar distribution. Burster locations that are compatible with the BATSE observations include an extended galactic halo, sampled to distances of a few hundred kiloparsecs (but less than 400 kpc to avoid an unobserved dipole moment in the direction of M31), or cosmological distances, for which the inhomogeneity arises from geometric effects at redshifts near unity (Harding 1994, Hurley 1994, Hakkila *et al.* 1994b). The perceived unnaturalness of the remaining galactic halo models (Hakkila *et al.* 1994b, Hakkila *et al.* 1994a) has led to a growing, though not universal, belief that GRBs are probably extragalactic.

A very large number of burst models have been proposed (Nemiroff 1994). Among the cosmological models, some involve, for example, AGN or cosmic strings and so do not predict local bursts. Most, however, predict occasional bursts in our Galaxy, and it is with these models that we are concerned. For example, a particularly popular model is the merger of two neutron stars after gravitational-radiation-driven inspiral of a binary system (Paczyński 1993). The model is attractive because such events are known to occur (Taylor and Weisberg 1989), and they release adequate total energy: $10^{53} - 10^{54}$ erg. As only three close double neutron star binaries are known in our galaxy, their birthrate (and hence inspiral rate) is difficult to estimate (van den Heuvel 1994), but is probably about $10^{-5}$ yr$^{-1}$. This rate is in reasonable agreement with a model independent estimate, obtained by assuming that GRB activity traces mass (Paczyński 1993): $2 \times 10^{-6}$ yr$^{-1}$. This estimate is at least roughly correct whether GRBs are actually double neutron star inspirals or are some more exotic phenomenon, subject only to the assumptions that they occur in normal galaxies and the bursts observed by BATSE are at cosmological distances.

If bursts are assumed to have a common peak luminosity, then the observed number-flux relationship can be used to estimate (Fenimore *et al.* 1993) that the faintest BATSE bursts are at a redshift $z \sim 0.8$, giving a luminosity in the 30-2000 keV passband of $\sim 6 \times 10^{50}$ erg s$^{-1}$. The total energy in this band for a burst of typical duration is therefore $\sim 10^{52}$ erg. Measurements of GRB spectra well above an MeV are relatively rare, but observations by the EGRET instrument (Sommer *et al.* 1994) of the bright burst GRB of 1993 Jan 31 showed comparable energy flux above and below an MeV. (Beaming can decrease the required burst energy, but it increases the required event rate by the same factor and leaves the conclusions of this work unchanged.)

The radiation at the Earth due to a "cosmological" ($10^{52}$ erg) GRB occuring at the center of our Galaxy would be $10^6$ erg cm$^{-2}$. The effect of such a gamma-ray pulse on the Earth's atmosphere was examined 20 years ago by Ruderman (Ruderman 1974), who considered high energy bursts associated with supernova explosions. The hard x-rays and gamma rays are stopped in the lower stratosphere, 20–30 km, by Compton scattering, resulting in the ionization of molecular nitrogen and the formation of nitric oxide and related "odd-nitrogen" compounds (NO$_x$). The importance of this process arises from the fact that NO$_x$ is a catalyst for the destruction of ozone, accounting at present for about 45% of ozone removal from the stratosphere (National Resource Council 1982). Assuming an efficiency of one NO$_x$ per 35 eV in gamma rays (Turco *et al.* 1982), about $3 \times 10^{34}$ molecules of NO$_x$ would be produced, comparable to the annual natural production and about a third of the total stratospheric NO$_x$ burden (Turco 1985). This NO$_x$ would be globally distributed by horizontal transport, and could remain in the stratosphere for several years before being transported to the troposphere or mesosphere and destroyed, in the mean time decreasing the ozone column by about 5% (National Resource Council 1982). This is comparable to depletions observed after the eruption of Mount Pinatubo (Gleason *et al.* 1993) and after the largest solar particle events associated with flares (Heath, Krueger, and Crutzen 1977, Stephenson and Scourfield 1991), and somewhat smaller than the depletion believed to occur after the Tunguska meteor fall (Turco *et al.* 1982).

Of course, if burst sources are distributed in the Galaxy, then some will occur closer to the Earth. For example, PSR B1534+12 and its companion (Wolszczan 1991) will coalesce in $\sim 10^9$ yrs and are only about 0.5 kpc distant. A GRB at that distance would bathe the earth with $\sim 3.5 \times 10^8$ erg cm$^{-2}$, equivalent to a total of $10^4$ megatons of TNT, or roughly the total worldwide nuclear arsenal. (By comparison, the total chemical energy of the ozone layer is only 3000 megatons (Hampson 1974).) An event of this magnitude could essentially destroy the ozone layer for years, and could occur every few hundred million years.

A decrease in the amount of stratospheric ozone, even if not catastrophic, leads to an increase in the amount of erythemally active ultraviolet-B radiation (UVB, 290-320 nm) at the Earth's surface. In four years of observations in Toronto, as average ozone levels decreased UVB was observed to increase with an amplification factor between 1.1 and 1.3, depending on season (Kerr and McElroy 1993). This is in good agreement with simple models (National Resource Council 1982): a 50% reduction in ozone is expected to increase the intensity by a factor 2 at 305 nm and a factor 50 at the more biologically active 295 nm.

The effects of this increased UVB on the biosphere are extensive and complex (National Resource Council 1982). The very existence of the biosphere probably depends on the existence of the ozone layer (which would be only 3-4 mm thick at STP) (Hampson 1974). The most familiar consequences of UVB exposure in humans are sunburn, or erythema, and increased risk of melanoma and other cancers, but the ecologically most important effects probably arise lower on the food chain. UVB exposure inhibits growth of many plants, with a 50% ozone reduction approximately doubling the biocidal effects of UVB (National Resource Council 1975). Many key organisms in aquatic ecosystems live in a close balance with existing radiation levels, with UVB tolerances approximately equal to exposure (Calkins and Thordardottir 1980). While such organisms may adapt to a slowly increasing level of UVB, they may be unable to respond to a rapid change. Ozone depletion also changes the ratio of UVB to the longer wavelength UVA-PAR, depriving animals of a key cue used for UVB avoidance (Bothwell, Sherbot, and Pollock 1994). A moderate change in ultraviolet flux may thus be damaging to individuals and species already under stress, and a large change may have correspondingly dramatic, biosphere-wide effects.

Photochemical changes in the stratosphere may have other important effects as well. For example, oxidation of NO$_x$ into HNO$_3$ will lead to increased acidity of precipitation for several years following a nearby GRB. The $10^7$ tons of nitrogen in NO$_x$ produced by a GRB at the galactic center is half the annual anthropogenic release to the troposphere (Bricker and Rice 1993); a closer GRB could, through this mechanism, substantially increase surface water acidity. Increased mortality may be expected for many species, though the nitrogen may also act as a fertilizer, increasing productivity of some plant species.

Environmental radioactivity can also be expected to increase following a nearby GRB, with potentially detectable consequences. A number of GRBs have been observed above 10 MeV, and photons up to 26 GeV were detected from GRB 940217 (CGRO status report). This high energy flux will cause photonuclear reactions in the atmosphere: for example, atmospheric production of $^{14}$C by gamma rays with energy $\gtrsim$ 10 MeV has a yield of about $10^3$ atoms per erg (Lingenfelter and Ramety 1970), compared with the average cosmic-ray induced production rate (Lingenfelter and Ramety 1970) of $2.1 \pm 0.4$ at cm$^{-2}$ s$^{-1}$. A burst closer than a kpc could, in 10 s, produce as much $^{14}$C as 1000 yrs of cosmic rays. With a 5700 year halflife, such an overproduction is unlikely to be detectable today, but other radionuclides might preserve a record of galactic GRBs. $^{10}$Be, formed by spallation on O and N, has a halflife of 1.5 Myr, and marine sediments have a memory of $^{10}$Be production as long ago as 10–15 Myr, though with relatively poor temporal resolution (Morris 1991). Cosmic-ray induced production of radionuclides takes place preferentially at high latitudes, where charged particles penetrate deeply into the atmosphere, while photonuclear production will occur hemispherically. A radionuclide enhancement due to a GRB could thus in principal be distinguished from one due to an increased cosmic ray flux. The traces of a nearby GRB may also be detectable through the radionuclear record in meteorites or the lunar surface (Reedy, Arnold, and Lal 1983, Vogt, Herzog, and Reedy 1990), though careful analysis will be required to distinguish the effects of high energy gamma rays from cosmic rays.

The relationship of binary inspiral events to GRBs may be settled by detection with LIGO of a gravitational wave burst coincident with the gamma rays. Such a finding, or any other definite link between GRBs and systems found in our galaxy, will imply that the Earth has on occasion found itself uncomfortably close to a burst source: within 1 kpc every $\sim 10^8$ yrs. It is tempting to speculate that some of these bursts may have led to mass extinctions, which have occurred at similar intervals (Russell 1979), but a complete understanding of the effects of such a nearby burst will require understanding the changes in stratospheric chemistry and transport processes that occur during very large perturbations. If GRBs prove to be at cosmological distances, then the flashes BATSE observes from the furthest reaches of the Universe may have close ties to life here on Earth.

We thank Z. Arzoumanian, B. Paczyński, D. Chakrabarty, R. Dewey, D. Hogg, V. Kaspi, S. Kulkarni, P. Ray, and J. Taylor for discussions and comments.